\documentclass[aps,pra,reprint, amsmath, amssymb,superscriptaddress]{revtex4-1}
 
\usepackage{bm}
\usepackage[retainorgcmds]{IEEEtrantools}
\usepackage{graphicx}
\usepackage{mathrsfs}
\usepackage{amsmath}
\usepackage{amsfonts}
\usepackage{amssymb}
\usepackage{color}
\usepackage{times,txfonts}
\usepackage{nicefrac}
\usepackage{ragged2e}
\usepackage{tikz}

\usepackage[colorlinks=true,linkcolor=blue,urlcolor=blue,citecolor=blue,pdfusetitle]{hyperref}

\usepackage{blindtext}

\definecolor{cadmiumgreen}{HTML}{097969}

\newcommand{\blah}[1]{blah\\blah\\blah\\blah.}

\begin{document}

\title{Classical dissipative cost of quantum control
}

\date{\today}
\author{Anthony Kiely}
\email{anthony.kiely@ucd.ie}
\affiliation{School of Physics, University College Dublin, Belfield, Dublin 4, Ireland}
\affiliation{Centre for Quantum Engineering, Science, and Technology,
University College Dublin, Belfield, Dublin 4, Ireland}
\author{Steve Campbell}
\affiliation{School of Physics, University College Dublin, Belfield, Dublin 4, Ireland}
\affiliation{Centre for Quantum Engineering, Science, and Technology,
University College Dublin, Belfield, Dublin 4, Ireland}
\author{Gabriel T. Landi}
\email{gtlandi@gmail.com}
\affiliation{Instituto de F\'isica da Universidade de S\~ao Paulo,  05314-970 S\~ao Paulo, Brazil.}
\affiliation{School of Physics, Trinity College Dublin, College Green, Dublin 2, Ireland}

\begin{abstract}
Protocols for non-adiabatic quantum control often require the use of classical time varying fields. Assessing the thermodynamic cost of such protocols, however, is far from trivial. Here we study the irreversible entropy produced by the classical apparatus generating the control fields, thus providing a direct link between the cost of a  control protocol and dissipation. We focus, in particular, on the case of time-dependent magnetic fields and shortcuts to adiabaticity. Our results are showcased with two experimentally realisable case studies: the Landau-Zener model of a spin-1/2 particle in a magnetic field and an ion confined in a Penning trap.

\end{abstract}
\maketitle{}

\section{Introduction}
Achieving realisable, robust, and high efficacy control is a crucial challenge in the development of modern quantum technologies~\cite{Deutsch2020}. One approach is slow adiabatic driving, but such long process times are inevitably susceptible to environmental spoiling effects. Ideally, one aims to achieve high target state fidelities on timescales faster than the decoherence rate. However, ramping a system arbitrarily will lead to non-adiabatic transitions which will similarly spoil the performance. Thus, there has been significant efforts to overcome this problem, and a particularly fruitful approach is captured by shortcuts to adiabaticity (STA)~\cite{Guery-Odelin2019}. These techniques can achieve an effective adiabatic dynamics in arbitrarily short times by implementing a specially designed control Hamiltonian. The viability of such strategies has been demonstrated in various experimental setups~\cite{Bason2012,Zhang2013,Rohringer2015}.

There has been a great deal of interest in clearly establishing how the implementation of such controlled evolutions imply an unavoidable cost in terms of some expended resource~\cite{Auffeves_2021}. While intuitively one would expect that fast driving should incur a higher penalty, it is nevertheless far from clear how to properly gauge this, particularly from a thermodynamic viewpoint~\cite{KosloffEntropy}. Early work on STA attempted to address this question for a variety of setups~\cite{Chen2010En,Torrontegui2017,Tobalina2018}. In parallel, several measures of cost have been proposed such as the norm of the driving Hamiltonian~\cite{Demirplak2008,Santos2015,Zheng2016, CampbellPRL2017,Deffner_2021,Carolan2022}, the work fluctuations \cite{Funo2017}, and the excess work~\cite{Bravetti2017} amongst others~\cite{Herrera2014,Calzetta2018,Abah2017,Abah2018,Cakmak2019}. While many of these approaches focused on the primary system being controlled, Torrontegui {\it et al} highlighted the importance of also considering the controller in assessing the resource intensiveness of a given protocol~\cite{Torrontegui2017}. Their model was classical however, so system and controller were of comparable dimensions. 
In quantum control protocols this is never the case: Time-dependent Hamiltonians are implemented by classical fields, produced by macroscopic apparatuses. There is therefore a fundamental asymmetry, between the microscopic system, and the macroscopic controller.

In this work, we begin with this fundamental observation: \emph{quantum control protocols are always generated by classical apparatuses}. Hence from the second law of thermodynamics, they will always be accompanied by a cost~\cite{Takahashi2017,Boyd2018}, associated to the dissipation in said apparatus. We analyse the irreversible entropy production~\cite{Fermi1956,Esposito_2009,Jarzynski_2011,Seifert_2012,Landi2021} --- a thermodynamic measure of dissipation --- associated with the generation of this classical field. In contrast to previous works, where costs are largely defined in an ad-hoc manner (see Ref.~\cite{Guery-Odelin2019} for a discussion), our approach gives an unambiguous connection to a directly measurable physical quantity. This is based on recent results for the stochastic thermodynamics of circuit elements, and is thus robust and extendable to other settings. We establish that while the qualitative behaviour is inherently protocol and setup dependent, this approach nevertheless provides a concrete, and experimentally meaningful, notion of cost. 
We show that the entropy production consists of two complementary components, which scale differently with the ramp duration,  allowing to us identify an optimal driving time for which the classical entropy production is minimised.

\section{Irreversible entropy of the control protocol} 
To make our ideas concrete,  we will neglect the generation of any static fields which play no direct role in the control protocol, and rather focus on time-dependent magnetic fields.  The starting point of our analysis is rooted in the observation that time-dependent  fields for a designated protocol must be produced by classical circuit elements. The entropy production for such a setup is connected to Joule heating and the Johnson-Nyquist fluctuations~\cite{Landauer_1975}. More recently, it has been incorporated within the framework of classical stochastic thermodynamics,  leading to a robust formulation, applicable to generic circuit elements and architectures~\cite{Bruers_2007,Landi_2013,Freitas_2020}.

\begin{figure}[t]
\begin{center}
\includegraphics[angle=0,width=0.9\linewidth]{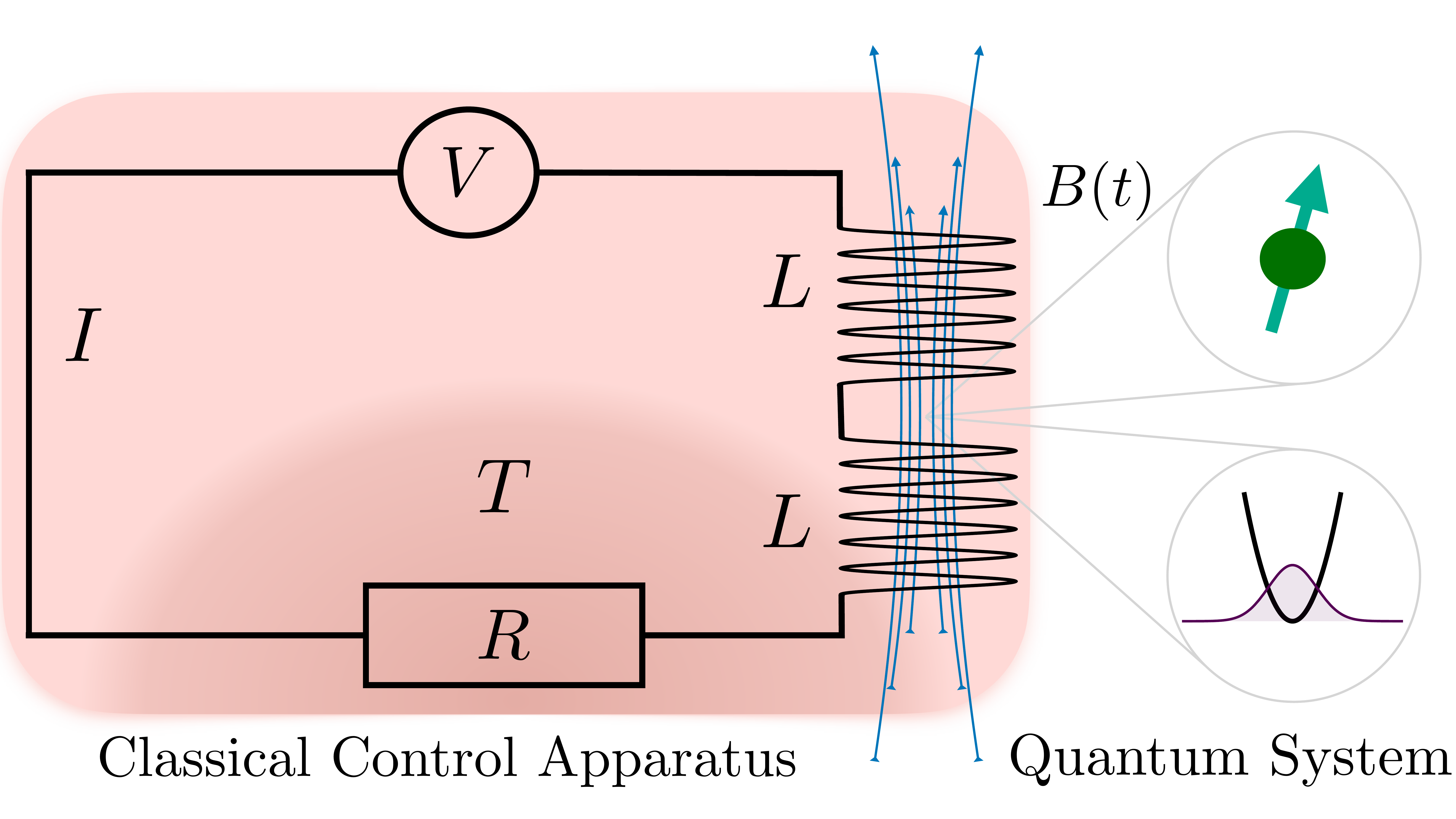} 
\end{center}
\caption{Schematic of a classical control device acting on the relevant quantum system of interest. In this setting, the classical device is an RL circuit at a temperature $T$ which generates an approximately uniform unidirectional magnetic field via a Helmholtz coil. The quantum system considered is either a single spin or an ion already in an electrostatic quadrupole potential i.e. a Penning trap.}  
\label{fig_diagram}
\end{figure}

With the basic setting established, a natural question is whether one can connect the entropy production with properties of the unitary driving. To answer this, we consider the case of magnetic fields generated by a Helmholtz coil, cf. Fig.~\ref{fig_diagram}. The field in the coil's axial direction will be given by $B(t)\!=\! A I(t)$, where $A$ is a constant depending on the coil geometry and $I(t)$ is the coil's current.


A given quantum control protocol determines the required field, $B(t)$, which in turn fixes  $I(t)$. To generate this current, we assume a direct drive scheme using a function generator with output voltage $V(t)$. Assuming negligible capacitance in the coil, the average current will be determined by $V(t)$ via the Langevin equation $2L \dot{I} + RI(t) \!=\! V(t) + \sqrt{2 R k_B T} \dot{\xi}(t)$, where $2L$ is the total inductance of the coils, $R$ the total electrical resistance, $k_B$ is Boltzmann's constant, $T$ is the temperature and $\dot{\xi}(t)$ is a Gaussian white noise presenting the Johnson-Nyquist fluctuations. Given a target average current $\langle I(t) \rangle$, we can use this to reverse engineer the required $V(t)$. The choice of control protocol thus ultimately determines the voltage that must be supplied to the function generator. 

We note that a classically fluctuating control field $B(t)$ will cause dephasing in the basis of the operator which it implements. The resulting small error in the fidelity can be minimised by an appropriate choice of STA control protocol, see e.g \cite{Ruschhaupt_2012}. 

In this setting entropy is constantly produced due to Joule heating in the resistor. The instantaneous entropy production rate, $\dot{\Sigma}$, can be computed using results from stochastic thermodynamics of electric circuits~\cite{Landauer_1975,Bruers_2007,Landi_2013,Freitas_2020} (see Appendix~\ref{appA} for details),  and is given by 
\begin{equation}
\label{dot_sigma}
    \dot{\Sigma} = \frac{R}{T} \langle I(t) \rangle^2 + \frac{R}{4L^2} \frac{(k_B T-2L\Delta_I^2)^2}{T \Delta_I^2},
\end{equation}
where $\Delta_I^2$ is the current variance. The first term is essentially the dissipated heat in the resistor, while the second provides an additional source of irreversibility, due to fluctuations. Typical quantum control protocols are designed to minimise said fluctuations, so one would naturally expect this contribution to be small. Moreover, as shown in Appendix~\ref{appA}, $\Delta_I^2$ evolves independently of the voltage $V(t)$, and has a steady-state value $\Delta_{I,{\rm ss}}^2 = k_B T/2L$, precisely cancelling the last term in~\eqref{dot_sigma}. Therefore if the circuit is allowed to stabilise before the protocol is initiated, the last term will remain zero throughout, irrespective of the choice of $V(t)$. Assuming this is the case and integrating over the duration of the protocol we find 
\begin{equation}\label{Sigma}
    \Sigma 
    =\chi \int_0^\tau dt  \left \langle B(t)\right \rangle^2,
\end{equation} where $\chi\!=\! R/(T A^2)$ is a constant depending on various fixed parameters of the circuit. Note that if the fluctuations are not negligible, Eq.~\eqref{Sigma} will be a lower bound instead, since the second term in Eq.~\eqref{dot_sigma} is always non-negative.

Equation~\eqref{Sigma} represents the irreversible entropy dissipated due to a magnetic field control protocol and comes with two important consequences: {\it (i)} the implementation of any protocol comes with an unavoidable entropic penalty due to the classical circuitry used to implement the control fields; and {\it (ii)} this entropy production is inherently related to the specific physical setup. Thus, Eq.~\eqref{Sigma} neatly demonstrates that while quantum control is never free, quantitatively examining the ``cost" of control can only be meaningfully done in a setting specific manner, where the details of the physical architecture dictates what sort of control fields are needed, while the functional time dependence of these fields are fixed by the control approach employed. We demonstrate these points in the following by examining two experimentally relevant and complementary case studies: the Landau-Zener model and an ion confined in a Penning trap.

\section{Case study: Landau-Zener model}
Consider a single spin-1/2 particle subject to a time-dependent magnetic field 
\begin{equation}\label{H_LZ}
    H_0(t) = \hbar\Delta\sigma_x + \hbar g(t) \sigma_z, 
\end{equation}
where $\sigma_i$ are the Pauli matrices. We assume the system is initialised in the ground-state of $H(0)$ with $g(0)\!=\!-g_0<0$, and the goal is to drive it in a finite time, $\tau$, to the ground-state of $H(\tau)$ with $g(\tau)\!=\!+g_0$, thus passing through an avoided crossing at $g=0$. The quantum adiabatic theorem establishes that for insufficiently slow protocols, excitations will occur and the target state will not be achieved with perfect fidelity. However, quantum control allows for this process to be achieved in arbitrarily short finite times.

\begin{figure*}
    \centering
    \includegraphics[width=\textwidth]{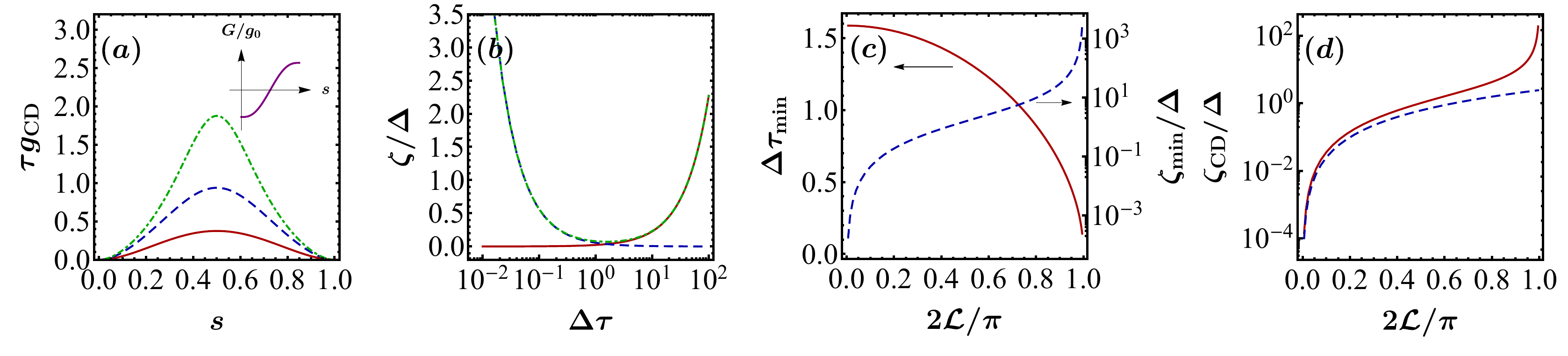}
    \caption{Entropy production in the Landau-Zener model with counterdiabatic driving.
    The system is driven under the Hamiltonian~\eqref{H_LZ} with a cubic polynomial pulse $G(s)$ shown in the inset of (a). 
    The resulting counterdiabatic driving $g_{\rm CD}$ is shown in (a) for $g_0/\Delta=\{0.2,0.5,1\}$(red solid, blue dashed, green dot-dashed).
    (b) The net entropy production of the $z$ driving $g(t)$, $\zeta_{\rm Z}$ (red solid), and the counterdiabatic driving $g_{\rm CD}$, $\zeta_{\rm CD}$ (blue dashed), as a function of the driving duration $\tau$, for $g_0/\Delta = 0.2$.
    The total entropy production $\zeta = \zeta_Z + \zeta_{\rm CD}$ is shown as a green dot-dashed line. 
    (c) Minimal entropy production $\zeta_{\rm min}$ and the operation time where this occurs $\tau_{\rm min}$ versus Bures distance $\mathcal{L}$. (d) Entropy production $\zeta_{\rm CD}$ versus Bures distance $\mathcal{L}$ for $\Delta \tau=1$, compared  with the lower bound in Eq. \eqref{eqlb} (blue dashed line).}
    \label{fig:LandauZener}
\end{figure*}

For such a system, one widely used control strategy is to introduce an additional counterdiabatic (CD) field~\cite{Demirplak2003,Berry2009,Chen2010LZ, Bason2012, Zhang2013}, $H_{\rm CD}(t) =  -\hbar g_{\rm CD}(t) \sigma_y$ with $g_{\rm CD}(t) = \frac{\Delta \dot{g}}{2(\Delta^2 + g(t)^2)}$.
Evolving the system with the total Hamiltonian $H\!=\!H_0+H_{\rm CD}$ leads to unit fidelity for any $\tau$ and any choice of $g(t)$. Intuitively, one expects that implementing this control will come at an unavoidable additional cost due to the additional field required~\cite{Abah2019}. To assess the entropy production associated with the generation of the classical control fields, we note that the control parameters are directly proportional to the magnetic field, $g(t)\!=\!\mu \langle B(t)\rangle/\hbar$, where $\mu$ is the magnetic moment. Evidently for this model there are three fields, in the $x,y,z$ directions, and we can assume that each is generated by an independent coil. However, since only the fields in the $z$ and $y$ directions vary in time, and thus are the only ones which play a role in the control protocol, in what follows we neglect the entropy production associated to the $\Delta \sigma_x$ field, which is simply a constant. Using Eq.~\eqref{Sigma}, the entropy production from the two dynamical fields can then be expressed as $\Sigma=\Sigma_Z+\Sigma_{\rm CD}$ where $\Sigma_i=\hbar^2\chi \zeta_i/\mu^2$ and
\begin{eqnarray}
  \zeta_Z&=&\tau\int_0^1 ds~G(s)^2, \label{entprodH0} \\
  \zeta_{\rm CD} &=&  \frac{1}{\tau} \int_0^1 ds~\frac{\Delta^2\dot{G}^2(s)}{4 [\Delta^2+G^2(s)]^2}, \label{entprodCD}
\end{eqnarray}
with $G(s)\!=\!g(s \tau)$ and $s\!=\!t/\tau$. 

From these expressions we find a clear connection between the classical entropy production with other established notions of cost. In particular, we see that $\Sigma_{\rm CD} \!\propto\! \int_0^\tau dt~||H_{\rm CD}(t)||^2$, and is therefore directly related to the operator norm based approaches studied in Refs.~\cite{Demirplak2008,Santos2015,Zheng2016}. It shows that the associated cost is not related to the average change in energy of the system, but rather to the integrated drive over the entire duration where $H_{\rm CD}$ is left on. Since the energy associated with $H_{\rm CD}$ is not strictly dissipated, it has been argued in Ref.~\cite{KosloffEntropy} that the invested resources for achieving control should be treated as a catalyst, rather than a cost per se. However, Eq.~\eqref{entprodCD} shows that even if one takes this view, there is still an unavoidable thermodynamic penalty to be paid due to the classical control fields.

It is further interesting to note that Eqs.~\eqref{entprodH0} and \eqref{entprodCD} exhibit different scalings as a function of $\tau$. The entropy production associated to the $H_{\rm CD}$ control driving field diverges for small $\tau$ and vanishes as the drive time increases. This is consistent with other analyses of the cost of quantum control, where the resources grow unboundedly as the protocol duration is reduced, while for drive times that approach the adiabatic limit there is no need for complex control protocols, and the associated costs vanish~\cite{CampbellPRL2017, Abah2019}. In contrast, the entropy production related to the bare Hamiltonian, Eq.~\eqref{entprodH0}, scales in the opposite manner. This leads to a non-trivial trade-off in determining the timescales when employing control is thermodynamically beneficial.

One can also address how $\zeta_Z$ and $\zeta_{\rm CD}$ depend on the distance between the initial ($g(0) = -g_0$) and the target ($g(\tau) = +g_0$) states. Clearly Eq.~\eqref{entprodH0} scales as $g_0^2$. As for Eq.~\eqref{entprodCD}, we show in Appendix~\ref{appB} that for \emph{any} control protocol $g(t)$, it satisfies the lower bound
\begin{equation}
    \zeta_{\rm CD} 
 \geq \mathcal{L}^2/\tau, 
 \label{eqlb}
\end{equation} 
where $\mathcal{L}\!=\!\arctan(g_0/\Delta)$ is the Bures distance between the initial and target states~\cite{Bures1969}. Equation~\eqref{eqlb} elegantly demonstrates the trade-off between speed and expended resources, inherent in achieving coherent control~\cite{CampbellPRL2017, Funo2017, Santos2015}, by connecting the entropy production with the geometry of the quantum states.

To demonstrate this point further, we consider a smooth driving ramp $ G(s)= g_0\big[20 s^3 - 30 s^4 + 12 s^5-1\big]$ (Fig.~\ref{fig:LandauZener}(a), inset), which avoids any discontinuities in the  field. The resulting counterdiabatic field is shown in the main panel of Fig.~\ref{fig:LandauZener}(a). The entropy production for both fields and the total entropy production as a function of ramp duration, $\tau$, are shown in Fig.~\eqref{fig:LandauZener}(b), where the $\zeta_{Z} \!\sim\! \tau$ and $\zeta_{\rm CD} \!\sim\! \tau^{-1}$ dependencies are visible. We see the total entropy production diverges in both limits corresponding to instantaneous and adiabatic protocols, cf. green dot-dashed curve in Fig.~\ref{fig:LandauZener}(b). Between these two extremes we find there exists an optimal driving time, $\tau_{\rm min}\!=\!\tau\sqrt{ \zeta_{\rm CD}/\zeta_Z}$, which corresponds to the minimal entropy production $\zeta_{\rm min}\!=\!2 \sqrt{\zeta_Z \zeta_{\rm CD}}$. In Fig.~\ref{fig:LandauZener}(c) we compare how this optimal drive time, and the associated total entropy production, depends on the Bures distance $\mathcal{L}$. We see that the optimal operation time $\tau_{\rm min}$  tends to $0$ as $\mathcal{L}\rightarrow \pi/2$. It is also apparent that transfer between distant states in Hilbert space requires higher entropy production.  Finally, Fig.~\ref{fig:LandauZener}(d) shows $\zeta_{\rm CD}$ as a function of the Bures distance for  $\Delta \tau=1$. The dashed lines show the bound~\eqref{eqlb} for comparison. In the limit of $\mathcal{L}\rightarrow 0$ there is no state transfer and hence no entropy production. Since a completely orthogonal state transfer in the Landau-Zener model requires infinite field strength, the entropy production diverges in the limit $\mathcal{L}\rightarrow \pi/2$. The behaviour of $\zeta_Z$~with~$\mathcal{L}$ is qualitatively similar.

\section{Case study: Ion in a Penning trap} 
As a second example, we consider an ion confined in a Penning trap~\cite{Penning1936,Brown1986}. In cylindrical coordinates, this consists of an electrostatic quadrupole potential $\phi(r,z)\!=\!\frac{m \omega_{z}^2}{4 q}\left(2z^2-r^2\right)$ and a uniform time varying magnetic field $\vec{B} \!=\!\left \langle B_z(t)\right \rangle \hat{z}$. The Hamiltonian for the motional state of the ion is~\cite{Kiely2015} $H(t) = - \frac{\hbar^2}{2m} \nabla^2 +
\frac{1}{2} m \omega_r^2(t) r^2 - \omega(t) L_z +\frac{1}{2} m \omega_z^2 z^2$
where $\omega_r^2\!=\!\omega^2-\omega_{z}^2/2$ is the radial trapping frequency, $\omega\!=\! q \left \langle B_z\right \rangle/(2m)$ is half the cyclotron frequency, $q$ is the charge, and $m$ is the mass. Note that the $z$-component of the angular momentum operator $L_z\!=\! \frac{\hbar}{i} \frac{\partial}{\partial \theta}$ is a conserved quantity and the $z$ coordinate decouples from $r$ and $\theta$.

With the ion initialised in the ground state, the goal is to vary the magnetic field $\left \langle B_z(t)\right \rangle$ to exactly reach the ground state for a different radial trapping frequency $\omega_r(\tau)$ in a finite time. Furthermore, here we consider an alternative control approach based on the formalism of Lewis-Riesenfeld invariants~\cite{Lewis1969,Kiely2015}. We design an auxiliary function $l(t)$, which is the characteristic radial length scale of the wavefunction. The aim is to change this length from $l(0)\!=\!l_0$ to $l(\tau)\!=\!l_\tau\!=\!l_0/\sqrt{c}$, which corresponds to $\omega_r(0)\!=\!c \omega_r(\tau)$. A value of $c>1$ therefore represents a compression of the trap, while $c<1$ is an expansion. The Bures distance between the two ground states is then given by $\mathcal{L}\!=\!\arccos \left( \frac{2 \sqrt{c}}{1+c} \right)$. Interestingly, a compression, $c\!>\!1$, results in the same Bures distances as an expansion, $c'\!=\!1/c\!<\!1$ i.e. $\mathcal{L}'\!=\!\mathcal{L}$. To ensure perfect fidelity and continuous fields, this auxiliary function must fulfil the boundary conditions $l(t_b)\!=\!l_{t_b}\!=\!\sqrt{\frac{\hbar}{2 m \omega_r\left(t_b\right)}}$ and $l^{(n)}(t_b)\!=\!0$ for $t_b\!=\!0,\tau$ and $n\!=\!1,2,3$.

The connection to the required magnetic field  can be succinctly expressed in terms of $l(t)\!=\!l_0 \lambda(s)$,  as $\left \langle B_z(s \tau)\right \rangle = \frac{\hbar}{q l_0^2} \frac{1}{\lambda(s)^2}
\sqrt{1 - \frac{\lambda(s)^3 \ddot{\lambda}(s)}{\eta^2} +
  \frac{\nu^2}{2-\nu^2} \lambda(s)^4},$
where $\eta\! =\!  \tau \omega_r(0)$ is a rescaled operation time and $\nu\! =\! \omega_z/\omega(0)$. Note that not all parameter combinations of $\eta$ and $\nu$ are possible in a physical setup (see Appendix~\ref{appC} for details). A possible choice of $\lambda$ which fulfils all the required boundary conditions is the minimal polynomial ansatz
$\lambda(s)=1+ 20 \alpha s^7-70 \alpha s^6+84 \alpha s^5 -35 \alpha s^4$,
where $\alpha\!=\!1-1/\sqrt{c}$.

The entropy production can be expressed as $\Sigma \!=\! \frac{\hbar^{3} \chi }{2 m q^2 l_0^6} \left(\zeta_d+\frac{\eta \nu^2}{2-\nu^2} \right)$ where we have
used $\tau\!=\!\hbar \eta /(2 m l_0^2)$ and defined the dynamical part of the entropy production to be
\begin{eqnarray}
\zeta_d &=& \int_0^1 
\left[\frac{\eta}{\lambda(s)^4} - \frac{ \ddot{\lambda}(s)}{\lambda(s) \eta} \right] ds. \label{dynent} 
\end{eqnarray}
Despite considering a different approach to achieving control, we find here important qualitative similarities with the Landau-Zener case. The entropy production also contains two terms: one which scales linearly with operation time and one which scales inversely with operation time. A fundamentally new feature of this example, however, is that now we can examine the differences between compression and expansion protocols. In fact, the two generally incur different costs, even if their Bures distances are equal.

\begin{figure}[t]
\begin{center}
\includegraphics[angle=0,width=0.49\linewidth]{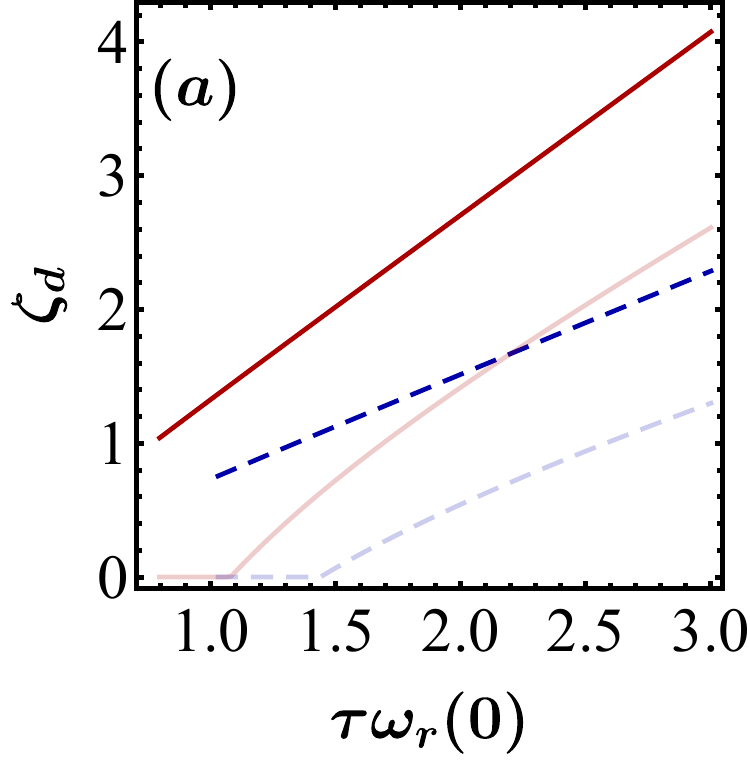} 
\includegraphics[angle=0,width=0.49\linewidth]{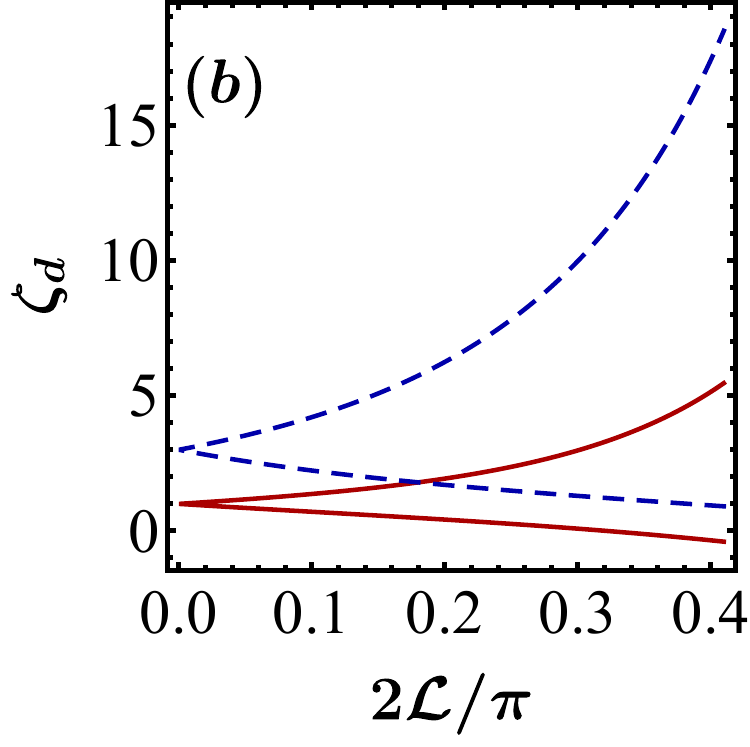}
\end{center}
\caption{Dynamical entropy production $\zeta_d$ versus (a) operation time $\eta$ for compression $c=4/3$ (red solid) and expansion $c=3/4$ (blue dashed) and (b) Bures distance $\mathcal{L}$ for $\eta=1,3$(red solid, blue dashed) with $\nu=1/2$. In (a) lower bounds are shown as faded lines. In (b) upper lines correspond to compression and lower lines to expansion.
\label{fig_penning}}  
\end{figure}

Fig.~\ref{fig_penning}(a) shows the entropy production against operation time for expansion and compression.  The results are only shown for times were the magnetic field remains real. For such operation times, it is clear that the linear term in $\zeta_d$ dominates. The faded lines show lower bounds for expansion $ \zeta_d \!\geq\! \max \left[0, c^2\eta  +\frac{84 |\alpha|}{5 \sqrt{5}\eta} \right]$ and compression $   \zeta_d \! \geq \!\max \left[0, \eta -\frac{84|\alpha| \sqrt{c}}{5 \sqrt{5}\eta} \right]$, developed in Appendix~\ref{appD}.  Fig.~\ref{fig_penning}(b) plots $\zeta_d$~vs.~$\mathcal{L}$. Note that $\zeta_d=\eta$ in the absence of any dynamics ($c=1$), which is clear from Eq.~\eqref{dynent}. For small values of $\mathcal{L}$, the entropy production is symmetric about $\eta$. However for increasing $\mathcal{L}$, the entropic cost of compression far exceeds that of expansion. This fundamental asymmetry is highly relevant to the compression/expansion strokes of a quantum heat engine \cite{Campo2014,Rossnagel2016}.

\section{Conclusions} 
We have examined the classical irreversible entropy produced by a control apparatus in achieving coherent quantum control. Our analysis starts from the simple observation that quantum control protocols are implemented through time dependent fields, generated by classical devices. This provided a natural framework, which can be directly related to currently established notions of cost, to quantitatively assess the thermodynamic penalty associated with quantum control. While there are several approaches to quantifying costs for quantum control~\cite{Guery-Odelin2019} (including the consideration of dissipation~\cite{Tobalina2018}) we establish a rigorous connection between the cost of quantum control and the entropy production which is the central quantity governing the second law of thermodynamics and encapsulates a more general notion of dissipation.

Our results therefore represent a crucial step towards solving the problem of how to assess the thermodynamic costs of a quantum control protocol. Our analysis clarified and elucidated several important results: {\it (i)} that quantum control comes with an inescapable cost that can be quantitatively examined by considering the associated macroscopic classical apparatuses that implements the fields; {\it (ii)} the framework presented is rooted in the stochastic thermodynamics of circuit elements, and therefore is naturally extendable to more complex settings; and finally {\it (iii)} that adiabatic and ultrafast protocols are thermodynamically inefficient as both lead to diverging entropy production. We were able to show that there can exist optimal driving timescales, where the competing effects between static and dynamic parts of the control protocol can reach a minimum. 

\acknowledgments
The authors acknowledge fruitful discussions with M. Mitchison, G. Guarnieri, T. Crilly, J. Xuereb, D. McGuire, J. Hackett and M. Doyle. AK and SC are supported by the Science Foundation Ireland Starting Investigator Research Grant ``SpeedDemon" No. 18/SIRG/5508. GTL  acknowledges the financial support of the Sa\~{o} Paulo Funding Agency FAPESP (Grant No. 2019/14072-0.) and the Brazilian funding agency CNPq (Grant No. INCT-IQ 246569/2014-0).


\appendix

\section{Instantaneous entropy production rate \label{appA}}

In this section we derive Eq.~\eqref{dot_sigma} of the main text, and also explain why the last term is often negligible. 
This equation is derived using the formalism of stochastic thermodynamics of electric circuit elements~\cite{Landauer_1975,Bruers_2007,Landi_2013,Freitas_2020}. 
Such systems are described by  Langevin equations~\cite{VanKampen2007}
\begin{equation}\label{SM_langevin}
    \dot{x}_i = f_i(\bm{x},t) + \sum_i B_{ij} \dot{\xi}_j(t), 
\end{equation}
where $\dot{\xi}_j$ are Gaussian white noises. 
Here $x_i$ represents a set of voltages and/or currents in the circuit, while $B_{ij}$ is a matrix characterising the Johnson noise in the circuit, due to thermal fluctuations. 
The voltages and currents are not all independent of each other, due to Kirchhoff's law. Techniques for obtaining the minimal set need to describe a circuit are discussed in~\cite{ElectricalNetworks}.
The corresponding probability distribution $\mathcal{P}(\bm{x},t)$ satisfies a Fokker-Planck equation
\begin{equation}
    \frac{\partial \mathcal{P}}{\partial t} = - \sum_j \frac{\partial}{\partial x_i} \Big[ f_i(\bm{x},t) \mathcal{P} \Big] + \sum_{i,j}D_{ij}  \frac{\partial^2 \mathcal{P}}{\partial x_i \partial x_j} := - \sum_i \frac{\partial g_i}{\partial x_i},
\end{equation}
where $D = \frac{1}{2}B B^{\rm T}$ is the diffusion matrix. The Fokker-Planck equation can be interpreted as a continuity equation in probability space, with the quantity $g_i(\bm{x},t)$ representing the probability current.  

To define the irreversible entropy production of the circuit dynamics, one must first establish which aspects of the dynamics are time reversible or not~\cite{Risken1989}. 
We define a variable $\epsilon_i$ such that $\epsilon_i = \pm 1$ whenever $x_i$ is even or odd with respect to time-reversal; voltages are even, while currents are odd. We then define~\cite{Landi_2013} 
\begin{IEEEeqnarray}{rCl}
f_i^{\rm irr}(\bm{x},t) &=& \frac{1}{2} \Big[ f_i(\bm{x},t) + \epsilon_i f(E\bm{x},t)\Big],
\\[0.2cm]
f_i^{\rm rev}(\bm{x},t) &=& \frac{1}{2} \Big[ f_i(\bm{x},t) - \epsilon_i f(E\bm{x},t)\Big],
\end{IEEEeqnarray}
where $E = {\rm diag}(\epsilon_1,\epsilon_2,\ldots)$. 
These represent the reversible and irreversible components of the Langevin equation~\eqref{SM_langevin}. 
With these definitions, we can now establish the corresponding components of the probability currents, by decomposing $g = g^{\rm irr} + g^{\rm rev}$, where 
\begin{IEEEeqnarray}{rCl}
    g_i^{\rm irr} &=& f_i^{\rm irr} \mathcal{P} - \sum_j D_{ij} \frac{\partial \mathcal{P}}{\partial x_j},
    \\[0.2cm]
    g_i^{\rm rev} &=& f_i^{\rm rev} \mathcal{P}.
\end{IEEEeqnarray}
The entropy production occurs due to the irreversible currents only $g^{\rm irr}$. In fact, the entropy production rate of the system is given, under rather general conditions, as~\cite{Qian_2001,Spinney_2012} 
\begin{equation}\label{SM_sigma_dot}
    \dot{\Sigma} = \sum_{i,j}\int \frac{d\bm{x}}{\mathcal{P}} (D^{-1})_{ij} g_i^{\rm irr} g_j^{\rm irr}.
\end{equation}
This formula is quite general, in that it applies to any kind of electric circuit subject to Johnson-Nyquist noise. It also applies to generic non-linear circuit models.

Next, we specialise it to the case of the Helmholtz coil model studied in the main text. 
The only variable in this case is the current $I(t)$ through the coils, which is odd under time reversal. 
The corresponding Langevin equation reads
\begin{equation}\label{SM_langevin_helmholtz}
    2L \dot{I} + R I = V(t) + \sqrt{2 R k_B T} \dot{\xi}(t).
\end{equation}
The corresponding irreversible probability current reads 
\begin{equation}
    g^{\rm irr} = - \frac{R}{2L} I \mathcal{P} - \frac{R k_B T}{L^2} \frac{\partial \mathcal{P}}{\partial I},
\end{equation}
while the diffusion coefficient reads $D = R k_B T/4L^2$. 
Plugging this in Eq.~\eqref{SM_sigma_dot} then yields 
\begin{equation}
    \dot{\Sigma} = \frac{R}{k_B T} \langle I(t)^2\rangle + \frac{R k_B T}{4L^2} \mathcal{F} - \frac{R}{L},
\end{equation}
where 
\begin{equation}
    \mathcal{F} = \int dI~\mathcal{P} \left(\frac{\partial}{\partial I} \ln \mathcal{P} \right)^2,
\end{equation}
is the Fisher information~\cite{Nicholson_2020}. 
To proceed, we use the fact that equation~\eqref{SM_langevin_helmholtz} is linear and therefore $\mathcal{P}$ will be  Gaussian. 
The corresponding Fisher information can then be computed exactly and reads $\mathcal{F} = 1/\Delta_I^2$, where $\Delta_I^2$ is the current variance.
Using also that $\langle I^2 \rangle = \langle I \rangle^2 + \Delta_I^2$, the entropy production rate finally simplifies to 
\begin{equation}\label{SM_dot_sigma}
    \dot{\Sigma} = \frac{R}{T} \langle I \rangle^2 + \frac{R}{4L^2} \frac{(k_B T - 2 L \Delta_I^2)^2}{T \Delta_I^2},
\end{equation}
which is Eq.~\eqref{dot_sigma} of the main text. 

Next, we show why the second term in this equation generally vanish. Using Eq.~\eqref{SM_langevin_helmholtz} one can develop an evolution equation for the variance $\Delta_I^2$. It reads 
\begin{equation}
    \frac{d}{dt} \Delta_I^2 = -(R/L) \Delta_I^2 + \frac{Rk_B T}{2L^2} .
\end{equation}
Crucially, one notices that this equation is independent of the applied voltage $V(t)$ (which therefore only affects the average current). 
Hence, irrespective of the quantum control protocol (determined by a specific shape of $V(t)$), the variance will tend to the steady-state $\Delta_{I,{\rm ss}}^2 = k_B T/2L$.
This cancels out the last term in Eq.~\eqref{dot_sigma} [or Eq.~\eqref{SM_dot_sigma}].
And since $\Delta_I^2$ is independent of $V(t)$, if the circuit is allowed to equilibrate before the control protocol starts, this term will remain zero throughout. 
Notice also that, if this is not the case, this term will simply add a corresponding positive quantity, so that Eq.~\eqref{Sigma} will be a lower bound to the actual entropy production.

\section{Lower bound for entropy production in the Landau Zener model \label{appB}}

Eq.~\eqref{entprodCD} for the entropy production associated to the counterdiabatic drive can be lower bounded as
 \begin{eqnarray}
 \zeta_{\rm CD} &=&  \int_0^\tau dt~g_{\rm CD}^2(t) \nonumber \\
 &\geq& \frac{1}{ \tau} \left(\int_0^\tau dt \, g_{\rm CD}(t) \right)^2 \nonumber \\
 &=& \frac{1}{ \tau} \left(\int_{-g_0}^{g_0} \frac{\Delta}{2(\Delta^2+g^2)} dg \right)^2 \nonumber \\
 &=& \frac{1}{ \tau} \arctan^2\left(\frac{g_0}{\Delta}\right) \nonumber \\
  &=& \mathcal{L}^2/ \tau,
 \end{eqnarray}
where we have used the Cauchy–Schwarz inequality $\left(\int_0^\tau f(t) dt\right)^2 \leq \tau \int_0^\tau f^2(t) dt$. 

\section{Penning trap: Physical Parameter limits \label{appC}}

To ensure that $\left \langle B_z\right \rangle \in \mathbb{R}$ for all times, the total time must fulfil the constraint that
\begin{eqnarray}
\eta \geq \max_{s \in [0,1]} \sqrt{\frac{\lambda(s)^3 \ddot{\lambda}(s)}{1+\nu^2 (2-\nu^2)^{-1} \lambda(s)^4}} .
\end{eqnarray}
A second constraint is that we wish to have a positive trapping potential i.e. $\omega_r^2>0$ at the start and end of the process. This limits the range to $0 < \nu < \sqrt{2} \, \mbox{min}\{1, \left \langle B_z(\tau)\right \rangle/\left \langle B_z(0)\right \rangle\}$. Note that the ratio of the initial and final fields is
\begin{eqnarray}
\frac{\left \langle B_z(\tau)\right \rangle}{\left \langle B_z(0)\right \rangle} =
\sqrt{c^2+(1-c^2)\frac{\nu^2}{2}},
\end{eqnarray}
which simplifies the final constraint to simply $0 < \nu < \sqrt{2} $.

\section{Lower bounds for entropy production in the Penning Trap \label{appD}}

Let us first focus on the case of expansion where $c<1$. We assume that $1=\lambda(0)\leq \lambda(s) \leq \lambda(1)=1/\sqrt{c} \, \forall s$. We also assume that the magnitude of the second derivative is upper bounded by $|\ddot{\lambda}|_{\rm max}$, which is dictated by the maximum magnetic field strength applied during the process. Since $\zeta_d \geq 0$, the lower bound reads
\begin{equation}
    \zeta_d \geq \max \left[0, c^2\eta   -\frac{|\ddot{\lambda}|_{max}}{\eta} \right].
\end{equation}
For our choice of $\lambda$ we have $|\ddot{\lambda}|_{\rm max}=\frac{84}{5 \sqrt{5}} |\alpha|$. For the case of compression $c>1$, we follow a similar logic and assume  $1/\sqrt{c}\leq \lambda(s) \leq 1$. The lower bound then reads
\begin{equation}
    \zeta_d \geq \max \left[0, \eta -\frac{|\ddot{\lambda}|_{max} \sqrt{c}}{\eta} \right].
\end{equation}


\bibliography{libtwo}

\end{document}